\documentclass[review,12pt]{elsarticle}
\makeatletter
\def\ps@pprintTitle{%
 \let\@oddhead\@empty
 \let\@evenhead\@empty
 \def\@oddfoot{\centerline{\thepage}}%
 \let\@evenfoot\@oddfoot}
\makeatother
\usepackage{hyperref}
\usepackage{graphicx} 
\usepackage{booktabs}
\usepackage{amsmath}
\usepackage{amsthm}
\usepackage{array}
\usepackage{epstopdf,subfigure}

\usepackage{caption} 
\newcommand{\tabincell}[2]{\begin{tabular}{@{}#1@{}}#2\end{tabular}}

\begin{document}
\renewcommand\arraystretch{0.6}
\begin{frontmatter}

\title{A new record of graph enumeration\\ enabled by parallel processing}

\author[mymainaddress,mysecondaryaddress]{Zhipeng Xu}
\ead{zhipeng.xu@stonybrook.edu}
\author[mysecondaryaddress]{Xiaolong Huang}
\ead{xiaolong.huang@stonybrook.edu}
\author[mythirdaddress]{Fabian Jimenez}
\ead{fjimenez@yachay.gob.ec}
\author[mysecondaryaddress]{Yuefan Deng\corref{mycorrespondingauthor}}
\ead{yuefan.deng@stonybrook.edu}
\cortext[mycorrespondingauthor]{Corresponding author}
\address[mymainaddress]{School of Data and Computer Science, Sun Yat-sen University, Guangzhou, Guangdong 510275, P.R. China}
\address[mysecondaryaddress]{Department of Applied Mathematics and Statistics, Stony Brook University, Stony Brook, NY 11794, USA}
\address[mythirdaddress]{Technologies, Empresa Pública Yachay, Centro de Emprendimiento Innopolis 100115, Ecuador}
\begin{abstract}
Using three supercomputers, we broke a record set in 2011, in the enumeration of non-isomorphic regular graphs by expanding the sequence of A006820 in Online Encyclopedia of Integer Sequences (OEIS), to achieve the number for 4-regular graphs of order 23 as 429,668,180,677,439, while discovering serval optimal regular graphs with minimum average shortest path lengths (ASPL) that can be used as interconnection networks for parallel computers. The number of 4-regular graphs and the optimal graphs, extremely time-consuming to calculate, result from a method we adapt from GENREG, a classical regular graph generator, to fit for supercomputers' strengths of using thousands of processor cores. 
\end{abstract}


\end{frontmatter}
\section{Introduction}
Analysis of regular graphs for their properties, including eigen-spectra and automorphisms, is a fertile field for discovering and applications in algebraic graph theory \cite{Godsil2001}. Yet, there are many unsolved problems, e.g., the Conway’s 99-graph problem \cite{conway} and the 57-regular Moore graph \cite{Hoffman1960}. For analysis of interconnection networks, regularity is essential for its direct and useful relationship to the complexity of network implementation, and as such, many regular graphs including the Peterson graph, hypercube graph, and their extensions \cite{Dasa,Ohringa,Ohring1993,Seo2011,Seo2017,Seo2008} are widely used to construct interconnection networks for parallel computers.
\par
For $3$-regular graphs of order $n$, Robinson and Wormald \cite{Robinson1983,Robinson1977} presented all counting results for $n\leq 40$, while pointing out that enumeration for unlabeled $k$-regular graphs with $k>3$ is an unsolved problem. Meringer \cite{Meringer1999} proposed a practical method to construct regular graphs without pairwise isomorphism checking but with a fast test for canonicity. Brankmann \cite{Brinkmann1996, Brinkmann2017} developed minibaum and snarkhunter for generating 3-regular graphs. “The House of Graphs” \cite{Brinkmann2013} and the OEIS \cite{inc2019line} databased the latest results for numbers of regular graphs. Kimberley \cite{A068934} contributed many results (A068934) to these databases by a package called GENREG developed by Meringer \cite{Meringer1999}. In addition to its challenges of pure mathematics, the graph counting problem is the root of topics in reliability, artificial intelligence, reasoning, statistical physics \cite{Vadhan2001}, life sciences, chemistry \cite{Meringer2010} and even the search for the origins of life \cite{Meringer2017}.
\par
GENREG, efficient for small-scale clusters due to its feature of task partition, approaches a hard wall of speedup for fine-grained partitioning on large-scale clusters, caused mainly by load imbalance. For obtaining larger graphs, we extend GENREG for distributed clusters by using the message passing interface (MPI) \cite{mpi2014}. Using the parallel GENREG we developed, we have obtained the following results:
\begin{enumerate}[1)]
\item filtered all 3-regular graphs up to order 32 with minimum ASPLs;
\item discovered thousands of 4-regular graphs of order 32 with minimum ASPLs;
\item generated the exact counts of 4-regular graphs of order 23 by using the three supercomputer clusters located in the US, China, and Ecuador.
\end{enumerate}
Among our results, the first and second have been applied to the interconnection network research \cite{Deng2019} for benchmark relationship of the graph ASPLs to network performance latencies. The third expands $n$ from 22 to 23, for the first time, in the sequence A006820 \cite{a006820} of OEIS, which is the number of connected 4-regular graphs of order $n$. Kimberley \cite{a006820} used GENREG to enumerate the 4-regular graphs for up to the order 22 in 2011 \cite{Larrion2016}. This record for $n=23$ remained unchallenged until our enumeration for $n=23$, enabled by our parallel computing implementation to advance it a step.
\section{The enumeration framework and results}
\subsection{The enumeration function}
For enumerating the regular graphs, published packages such as minibaum \cite{Brinkmann1996}, snarkhunter \cite{Brinkmann2017}, and GENREG have their own strengths and weaknesses, GENREG is more general in covering the graph degrees than minibaum and snarkhunter that only support 3-regular graphs.
\par
In our parallel computing framework, we designate one node as the master, whose task involves adaptive scheduling and dispatching, and the rest as a team of workers. When our program starts, the workers send a message to the master to request a task, and the workers continue the requests until the list of tasks exhausts. As usual, when the master sends a task to a worker, this task is marked as selected and becomes unavailable. At last, when the task pool empties, the master signals all workers to exit.
\par
Our dynamical scheduling strategy keeps cores in the cluster busy for useful tasks to allow us efficient search for graphs with specified parameters, e.g., diameters or eigenvalues, by inserting external serial programs. In addition to load balance, our parallel program reduces the communication cost to $N_{\text{task}}\times 2$, less requirement of bandwidth because the message itself is the message count. If we use a dedicated thread for task scheduling, the scalability and limit the maximum computer system can both shrink. Depending on the communication sub-system, the maximum scalable system our current approach can reach is approximately 3,000 cores, due to communication congestions, eventually. We may improve the scalability of our program by a multi-level scheduling; particularly, for the many-core systems.
\subsection{Search for a regular graph with minimal ASPL}
In the interconnection networks of supercomputers and data centers, regularity is a very significant feature because it is related to the complexity of the network configuration. For the topologies of regular graphs applied to the interconnection networks, it is highly desirable to obtain graphs with minimal ASPLs because they help reduce communication latencies. Let $d(i,j)$ be the distance between vertex $i$ and $j$, the ASPL is calculated as follows,
\begin{equation}
 \mathrm{ASPL}=\frac{1}{N(N-1)}\sum d(i,j)
\end{equation}
Cerf \cite{Cerf1975} calculated and proved lower bound of ASPL, hence the goal of optimal graph is to find graphs with minimum ASPLs. Usually and thus far, random or heuristic methods or intuitions were resorted to searching for graphs with such desired properties for large networks; Graph Golf \cite{golf}, a competition of searching for graphs with the minimal diameters and ASPLs, generated many graphs that are very commonly, but asymmetrically. However, we still hope to discover graphs with all desired properties including minimum diameters, ASPLs, symmetry, and robustness but with one disadvantage: smaller graphs. However, these graphs can be adopted in smaller clusters or multiple modules of these clusters or system-on-chip.
\begin{figure}[htbp!]
\centering
\subfigure[(32,3)-optimal]{
\includegraphics[width=6cm]{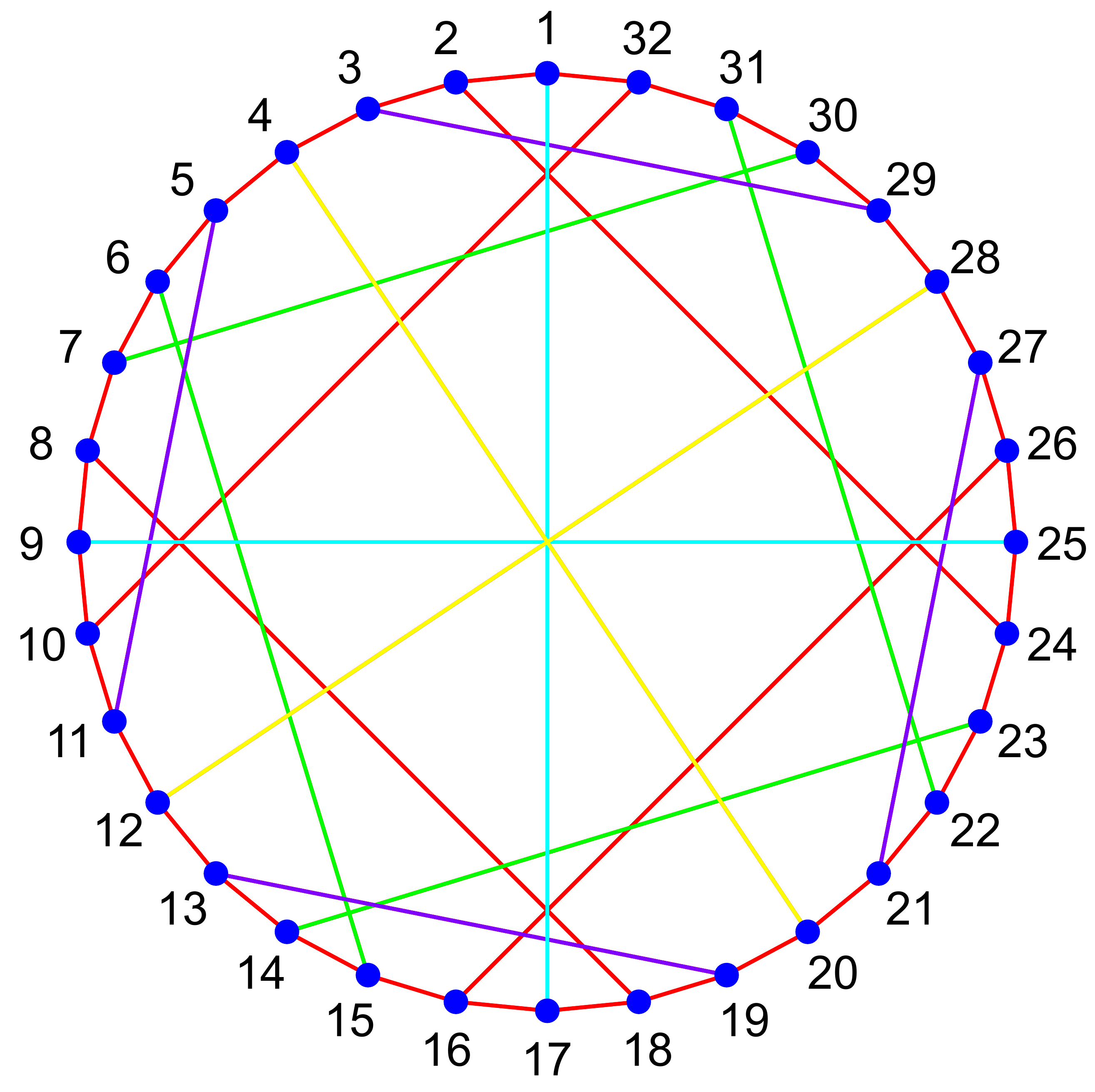}
}
\subfigure[(32,4)-optimal]{
\includegraphics[width=6cm]{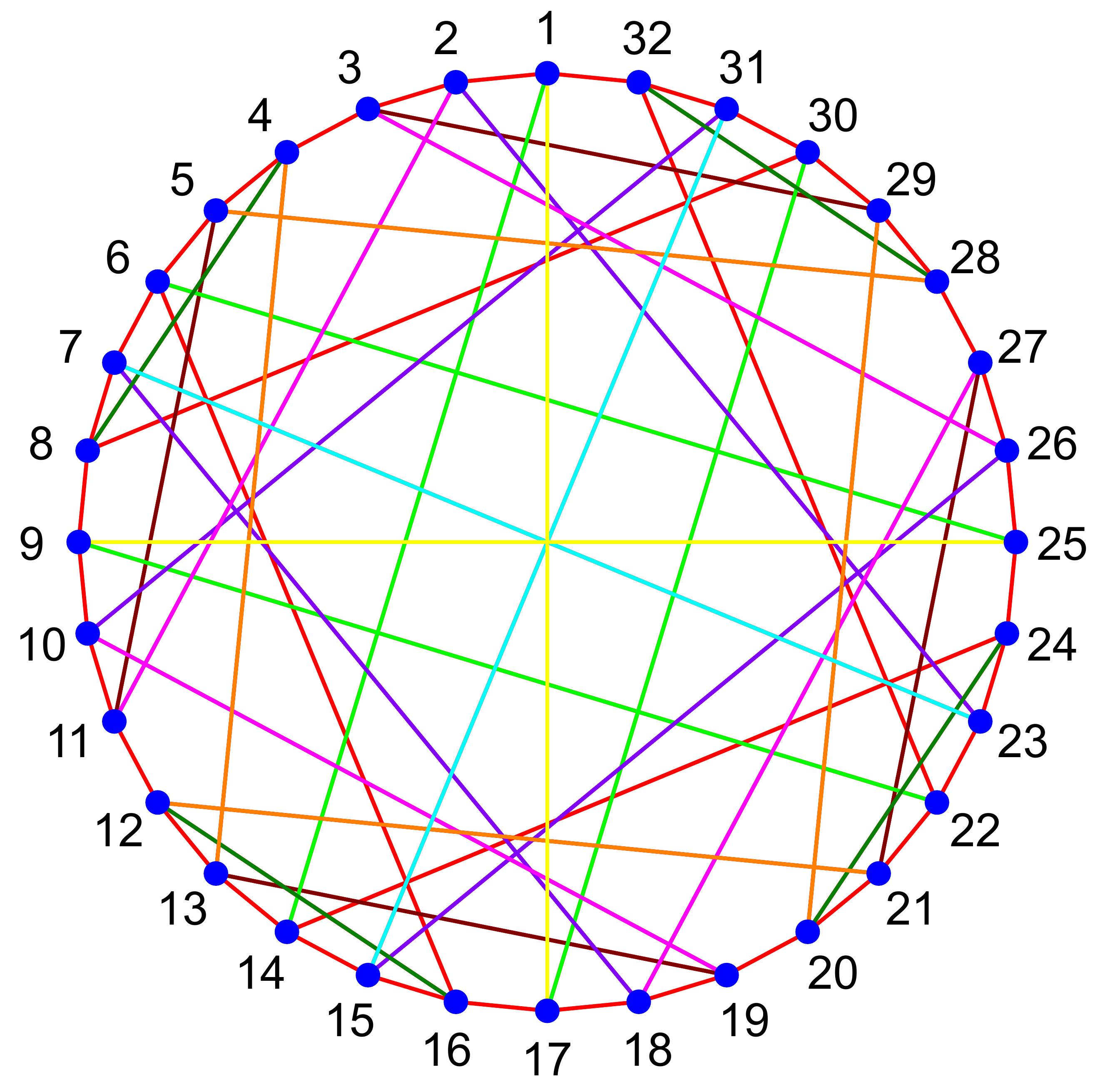}
}
	\caption{The optimal graphs used in, for example, a Beowulf cluster \cite{Deng2019}.}\label{32optimal}
\end{figure}
\par
Using this framework, we decomposed the search into 200,000 sub-tasks and completed the exhaustive search of all of 3-regular graphs of order 32 with diameter 4 and minimal ASPL. The program running on the Sunway-Bluelight supercomputer \cite{sunway} with 80,000 cores for 72 hours and discovered 56 graphs with the minimal ASPLs after exhausting all 18,941,522,184,590 possible graphs predicted by Robinson et al. [10] who only enumerated them without finding the graphs with desired properties including minimum diameters and ASPLs. Deng et al. \cite{Deng2019} applied one (Fig. \ref{32optimal} (a)) of these graphs to construct a Beowulf \cite{BE2003} cluster, and their benchmark results show that the graphs with the minimal ASPL outpeform other classical topologies. 
\par
\begin{figure}[htbp!]
	\centering
	\includegraphics[width=8cm]{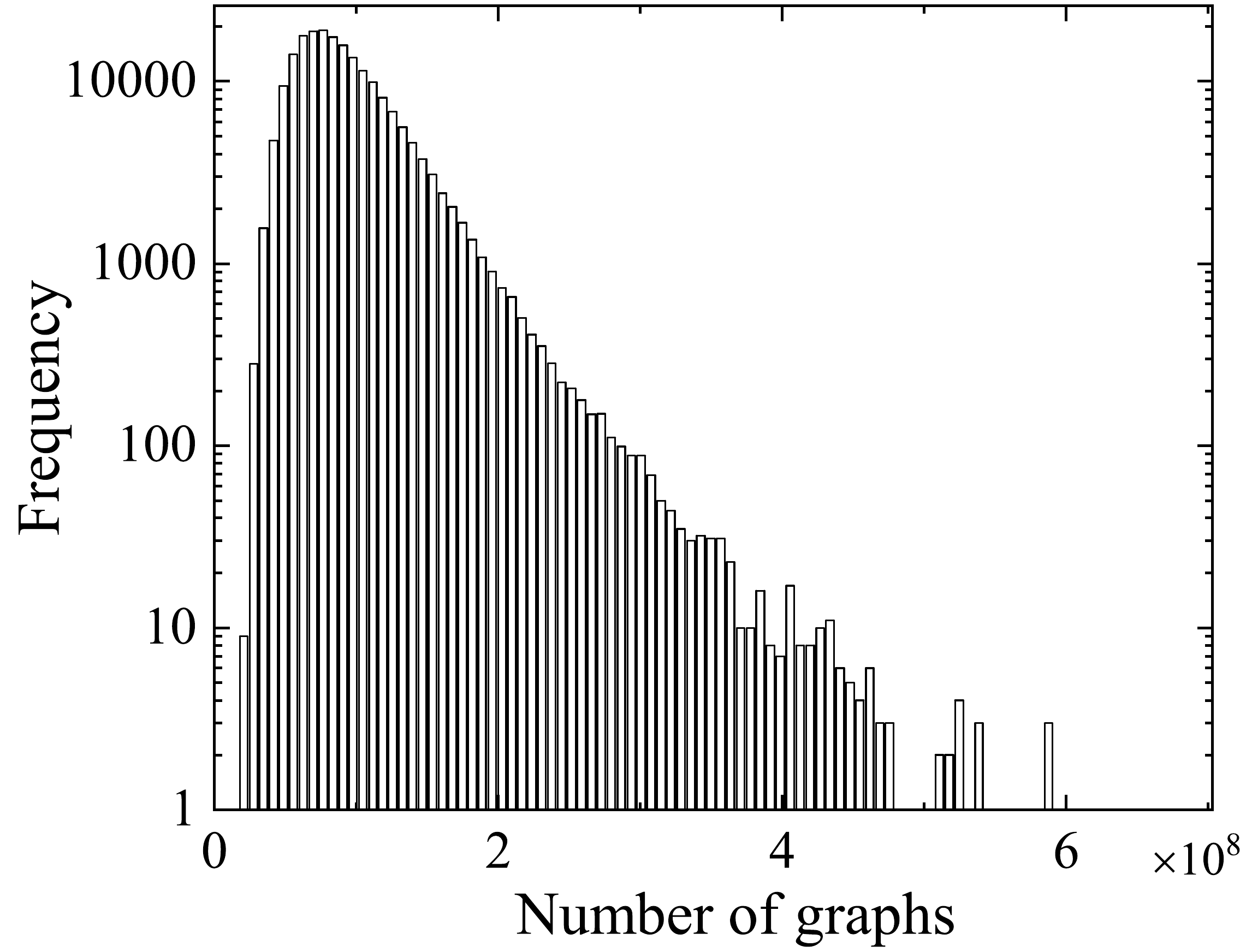}
	\caption{Number of graphs and frequency.}\label{32k3dist}
\end{figure}

Fig. \ref{32k3dist} shows the distribution of generated graph numbers for all cases, which look like Gaussian distribution with the long tail containing hundreds of millions of graphs and the frequency fluctuates by 4 orders of magnitude. Clearly, the tasks of enumerations vary greatly from graph to graph, and our dynamic task scheduling has eliminated substantial waiting time due to load imbalance.
\subsection{Graph counting for (23,4)-regular graphs}
When we search for the minimum-ASPL graphs among all possible 4-regular graphs of order 32, there is no enumeration result for this scale when $girth\leq 5$. We can still use our software to search for regular graphs with minimum ASPL, as shown in Fig. \ref{32optimal} (b), without confirmation of exhaustiveness. Therefore, we verified the results in Tab. \ref{result1} after decomposing the problem into 50,000 sub-tasks and setting the split-level as 12. We manage to increase the $n$ of the sequence A006820 from 22 to 23 and confirmed the number of 4-regular graphs of order 23 as 429,668,180,677,439 by using our parallel GENREG with the same parameters.
\begin{table}[]
\centering
\caption{OEIS A006820}\label{result1}
\begin{tabular}{@{}rr@{}}
\toprule
Order & Quartics            \\ \midrule
5     & 1                   \\
6     & 1                   \\
7     & 2                   \\
8     & 6                   \\
9     & 16                  \\
10    & 59                  \\
11    & 265                 \\
12    & 1,544               \\
13    & 10,778              \\
14    & 88,168              \\
15    & 805,491             \\
16    & 8,037,418           \\
17    & 86,221,634          \\
18    & 985,870,522         \\
19    & 11,946,487,647      \\
20    & 152,808,063,181     \\
21    & 2,056,692,014,474   \\
22    & 28,566,273,166,527  \\
\bf{23}    & \bf{429,668,180,677,439} \\ \bottomrule
\end{tabular}
\end{table}
\par
Our work to obtain the new enumeration for $n=23$ is estimated to cost nearly 100 core-years. We ganged three supercomputers, the SeaWulf at Stony Brook University, while the Tianhe-1 with Intel Xeon X5670 processors and the IBM Quinde with Power8 processors can process 113,000 and 56,469 graphs per second per core, respectively. As shown in Tab. \ref{clustes}, the Seawulf with Intel Xeon Gold 6148 processors has the highest efficiency for searching 178,000 graphs per second per core, while the Tianhe-1 with Intel Xeon X5670 processors and the IBM Quinde with Power8 processors contributed the rest.
\par
In fact, the result of graph counting for (23,4)-regular graphs was obtained by the strategic and opportunistic use of the fragmented and shared computing resources. In addition to the policy of fair and efficient share for most supercomputers, the Tianhe-1 would terminate tasks that last long for many cores because of maintenance, which prevents us from running long tasks. The external scheduling system we developed helps overcome the limitations of computing resources while facilitating optimal utilization of the occasionally available cores.
\begin{table}[]
\centering
\caption{Cost on three clusters}\label{clustes}
\begin{tabular}{@{}lrrr@{}}
\toprule
Cluster  & \tabincell{r}{Total processed\\($10^{12}$ graphs)}& \tabincell{r}{Computing time\\ (core years)} & \tabincell{r}{Core speed\\($10^3$ graphs/s) }     \\ \midrule
Seawulf  & $371.13$ & 66.12   & 178 \\
Tianhe-1 & $69.51$ & 19.53   & 113\\
IBM Quinde  & $23.60$ & 13.25   & 56 \\ \bottomrule
\end{tabular}
\end{table}
\section{Conclusions}
Our parallel method adapting the GENREG enables us to complete the research and enumeration on systems of 3000 processor cores. For the first time, using this new approach, we discovered several optimal graphs of order 32 and found the enumeration count for 4-regular graphs of order 23, gaining confidence in the graph theory community that high-performance computing can help solve otherwise intractable problems.
\section*{Acknowledgements}
Z. Xu is supported by the special project high performance computing of National Key Research and Development Program 2016YFB0200604. We thank the National Supercomputing Centers in Jinan and Changsha in China, for computing resources and, M. Meringer of German Aerospace Center, for technical support and of GENREG and beneficial suggestions of this manuscript via e-mails.
\bibliography{../Reference/Ref.bib}

\end{document}